\documentclass[aps,prl,amsmath,amssymb,nofootinbib,twocolumn]{revtex4}
\usepackage{amssymb}
\usepackage{color}
\usepackage{graphicx}
\usepackage{subfigure}
\usepackage{epsfig,amssymb,amsmath}
\usepackage[linkcolor=red]{hyperref}

\usepackage{color} 

\def\>{\rangle}
\def\<{\langle}
 \def\togli#1{}

\def\labell#1{\label{#1}}
\def\section#1{{\par\em #1:--- }}

\begin{document}


\title{Schr\"odinger cats and quantum complementarity} \author{Lorenzo Maccone}
\affiliation{ \vbox{Dip.~Fisica, Univ.~of Pavia, INFN Sez.~Pavia, via Bassi 6, I-27100
    Pavia, Italy} }

\begin{abstract}
  Complementarity tells us we cannot know precisely the values of all
  the properties of a quantum object at the same time: the precise
  determination of one property implies that the value of some other
  (complementary) property is undefined. E.g.~the precise knowledge of
  the position of a particle implies that its momentum is
  undefined. Here we show that a Schr\"odinger cat has a well defined
  value of a property that is complementary to its ``being dead or
  alive'' property. Then, thanks to complementarity, it has an
  undefined value of the property ``being dead or alive''.

  In other words, the cat paradox is explained through quantum
  complementarity: of its many complementary properties, any quantum
  system, such as a cat, can have a well defined value only of one at
  a time. Schr\"odinger's cat has a definite value of a property which
  is complementary to ``being dead or alive'', so it is neither dead
  nor alive. Figuratively one can say it is both dead {\em and}
  alive. While this interpretation only uses textbook concepts (the
  Copenhagen interpretation), apparently it has never explicitly
  appeared in the literature. We detail how to build an Arduino based
  simulation of Schr\"odinger's experiment based on these concepts for
  science outreach events.
\end{abstract}
\pacs{}
\maketitle

The Schr\"odinger's cat argument was published by Schr\"odinger's in
\cite{cat} and devised during a discussion with Einstein. It was a
provocation to show that quantum effects cannot be naively hidden in
the microscopic realm, but through linearity (and entanglement) they
can affect also macroscopic systems such as cats, giving rise, in
Schr\"odinger's words, to ``ridiculous cases''. While quantum
mechanics is by now well established, the debate on its interpretation
is still very much open: a huge number of competing interpretations
vie for an {\em explanation} of the quantum formulas. The
Schr\"odinger cat is a perfect testbed for such explanations
\cite{cattv}.

In this paper, we give a very simple interpretation to the cat, the
``complementarity interpretation'', based on textbook concepts, namely
on the Copenhagen interpretation. We then show how to use it to create
a simulated experiment which is effective to communicate to the
general public concepts such as the superposition principle and
quantum complementarity. Even though our explanation could have been
proposed by Bohr himself (who famously championed quantum
complementarity), Bohr never explicitly addressed the cat
\cite{arkady}. In his discussions with Schr\"odinger \cite{bohr}, he
did address classicality, but only limiting himself to the measurement
apparatuses, which is not what Schr\"odinger's argument
embodies. Surprisingly, to our knowledge the proposal presented in
this paper (Schr\"odinger's cat through complementarity) has not
appeared in the English literature previously \cite{lorenzoagiati}.

The outline of the paper follows. We start by explaining quantum
complementarity and how it is embodied in quantum mechanics. We then
recall Schr\"odinger's argument and show how quantum complementarity
can be used to make sense of the superposition. We conclude detailing
how one can construct a simulation of Schr\"odinger's experiment in a
cardboard box controlled through an Arduino microcontroller with a
simulated cat.

\section{Quantum complementarity}
All systems are described by a collection of their properties. For
example, the state of a featureless particle is described by two
properties, its position and its momentum. A spin 1/2 is described by
the components $J_x,J_y,J_z$ of its angular momentum. A cat is
described by a myriad of complicated properties (color, weight,
furriness, position, etc.). One of these is its ``being dead or
alive'': it can have two distinct values `dead' and `alive' (we will
neglect intermediate `moribund' values).

Properties in quantum mechanics are described by observables, such as
the position observable $X$. The states of quantum systems are
described by vectors $|\psi\>$, and the state of systems which have a
definite value of a property are eigenstates $|x\>$ of the
observable. The corresponding eigenvalue $x$ is the value of the
property in such state. States are vectors because of the
superposition principle: any state is a linear combination of
eigenvectors of some observable. Also any eigenstate of an observable
is a linear combination of eigenstates of other observables. The Born
rule tells us that, if we prepare a system in a (nontrivial)
superposition of eigenstates of the observable we are measuring, we
will get a probabilistic answer: that observable property is not well
defined in that system state. (The probabilities are given by the
square moduli of the linear combination coefficients. We discuss below
whether such probabilities are due to ignorance of something, or to
the fact that this something is undefined.)

Complementarity, hence, follows by joining superposition with the Born
rule. Consider two properties: observable $\hat H$ and observable
$\hat S$. Suppose that the $\hat H$ eigenstates are all nontrivial
superpositions of $\hat S$ eigenstates and that the state of the
system is in an eigenvector of the $\hat S$ observable. Then property
$\hat S$ will have a definite value, but $\hat H$ will
not. Complementarity \cite{bohrcom1,bohrcom} can be loosely stated as:
{\em ``if the value of one property is determined exactly, the values
  of some other properties will be undefined''}. E.g.~if the position
of the particle is exactly known, its momentum is completely
undefined: position eigenvectors are equal superpositions of {\em all}
momentum eigenvectors\footnote{Similarly, if the $z$ component $J_z$
  of the angular momentum of a spin $\tfrac12$ particle is exactly
  known, the $x$ and $y$ components are undefined: the $J_z=+\tfrac12$
  eigenstate $|J_z^{+}\>$ is an equally weighted superposition
  $|J_z^{+}\>=(|J_x^{+}\>+|J_x^{-}\>)/\sqrt{2}$ of the eigenstates
  relative to both possible values $\pm\tfrac12$ of the $x$ component
  $J_x$ (and similarly for $J_y$).}.  
Complementarity is usually vaguely defined in textbooks (if at
all), but it is just an aspect of the superposition principle (which is
similarly vaguely defined \cite{dirac}).
\section{The cat argument} Schr\"odinger suggests closing a cat in a
perfectly isolated box, which contains an ``infernal device'' which
opens a poison vial if an atom decays. He then suggest to use an atom
that has a half-life of one hour and to wait for one hour. What will
happen? Of course, the atom has probability one half of decaying, so
the cat has probability one half of dying. At first sight, this is not
paradoxical at all.

The paradox emerges if one analyzes more carefully the predictions of
quantum mechanics: since the atom is a quantum system, it evolves
through the Schr\"odinger equation which describes a deterministic
evolution: after one hour the atom has an equal probability amplitude
of being decayed or nondecayed. Namely, the state of the atom is an
equally weighted superposition:
$(|decayed\>+e^{i\varphi}|non\ decayed\>)/\sqrt{2}$ where $\varphi$ is
a phase that depends on the details of the evolution (we will choose
$\varphi=0$ for simplicity). Since the box is perfectly isolated, the
evolution of the whole box can be described through the Schr\"odinger
equation\footnote{We emphasize that, from the point of view of anyone
  outside the box, the evolution of the cat is {\it not} a
  measurement. Indeed, a perfectly isolated evolution cannot be seen
  as a measurement since the bare minimum that any quantum measurement
  must satisfy is that it must provide an {\it outcome}. An isolated
  evolution cannot provide any outcome by definition. Moreover, the
  quantum evolution postulate asserts that isolated systems evolve
  according to the Schr\"odinger equation. In this paper we will not
  be concerned of the point of view of the cat
  \cite{wignerfriend1,wignerfriend2}.}, and the cat (through the above
device) inherits the properties of the atom. Namely, the box prepares
the cat in a state
\begin{align}
  \tfrac1{\sqrt{2}}(|dead\>+|alive\>)
\labell{cats}\;,
\end{align}
where for simplicity of notation, the two kets $|dead\>$ and
$|alive\>$ represent the (entangled) state of all the degrees of
freedom in the box: all the atoms and photons that compose the cat,
its fleas, the poison vial, the radioactive atom, the molecules of air
in the cat's lungs and in the rest of the box and so on\footnote{While
  the nomenclature ``quantum entanglement'' was coined by
  Schr\"odinger in the cat paper \cite{cat} (and was discovered by
  Einstein, Podolsky, Rosen \cite{epr}), it is not mentioned in the
  exposition of the cat and appears much later in the paper.
  Entanglement is not really necessary to understand the cat argument,
  unless one wants to discuss separately the single degrees of freedom
  of the contents of the box.}. This state is a superposed state
(``smeared psi-function'', in the words of Schr\"odinger), which is a
situation physically {\em distinguishable} from the situation in which
the box contains a cat which is dead or alive with probability one
half. This last situation is described by a mixed state of the form:
\begin{align}
\tfrac12(|dead\>\<dead|+|alive\>\<alive|)
\labell{rhocat}\;.
\end{align}
The two states \eqref{cats} and \eqref{rhocat} are distinguishable
only if one looks at a property that is complementary to ``being dead
or alive'', as discussed below. If, instead we {\em only} consider the
property of ``being dead or alive'': we cannot distinguish them, they
both describe a box containing a cat that is dead or alive with
probability one half.

Consider two complementary observables: the first observable ``being
dead or alive'' $\hat H$ has the two orthogonal states $|dead\>$ and
$|alive\>$ as eigenstates, corresponding to the two possible values of
the cat's health; the second observable $\hat S$ is defined as the one
with eigenstates $|+\>=(|dead\>+|alive\>)/\sqrt{2}$ and
$|-\>=(|dead\>-|alive\>)/\sqrt{2}$ and eigenvalues $+1$ and $-1$
respectively. For lack of better words, we will call this property
``plus or minus''. The observable $\hat S$ can be measured by using
some quantum interferometric experiment \cite{wolfgang}, where the
$|+\>$ and $|-\>$ states refer to constructive and destructive
interference respectively. Of course, the apparatus which measures
such an observable will be insanely impractical (understatement)
\cite{wolfgang}, but there is no in-principle reason why it cannot be
built. The $\hat S$ property and the $\hat H$ property are
complementary, since the eigenstates of one are superpositions of the
eigenstates of the other, and viceversa.

Since the cat is in the state \eqref{cats}, the measurement of the
``plus or minus'' $\hat S$ observable will have outcome $+1$ with
certainty.  Thus, the cat possesses a definite value (plus) of the
$\hat S$ ``plus or minus'' property, which is complementary to
$\hat H$ ``being dead or alive''. Thus it is neither dead nor
alive. As a figure of speech, it is customary to say that it is ``dead
AND alive at the same time'' (since it's neither).

This is the paradoxical situation: we have experience of dead cats or
of alive cats. We can even easily think of cats in a box of which we
do not know if they are dead or alive, a situation described by
\eqref{rhocat}. However, we cannot even imagine a quantum-superposed
(``smeared'') cat that is neither dead nor alive, described by the
state \eqref{cats}. It would not even remotely look like a cat: it
would look like some interference pattern in an enormously complicated
interferometer (even though it {\em is} a cat: the interferometer is
showing a strange property {\em of the cat} that is complementary to
its usual properties). It is so inconceivable to think of such a
``cat'' that we do not have appropriate words to describe the
situation (except in the language of mathematics) and we resort to
somewhat inappropriate ``dead AND alive'' statements. The reason we do
not have experience of such cats is due to the incredible complexity
of the experiment necessary to measure the $\hat S$ property in a
cat. This, in turn, is due to decoherence: the almost unavoidable
interaction with the environment of macroscopic systems such as a cat
means that some of the cat's properties (such as $\hat H$) are
evident, whereas others (such as $\hat S$) are not. This selection of
evident properties (technically, einselection \cite{wojciech}) is due
to the fact that local interactions with superpositions of states
localized in macroscopically different positions (such as a
superposition of a dead and alive cat) will decohere rapidly: the
environment becomes immediately entangled with the system. Instead,
localized states such as a dead or an alive cat will not decohere at
all. This is the modern view of the emergence of classicality from a
quantum world. In other words, the reason why we never see superposed
cats is, according to quantum mechanics, only due to the practical
difficulty in isolating a cat from its environment and in carrying out
the appropriate interferometric experiment, and not because of any
fundamental reason: sophisticated enough experiments can achieve it
in practice \cite{catexp1,catexp2}, although it would be vastly
impractical on the scale of a cat \cite{wolfgang}.

In summary, using quantum complementarity, the cat paradox is readily
explained: There exist a property $\hat S$ ``plus or minus'' that is
complementary to $\hat H$ ``dead or alive''. Since the cat has a
definite value of the ``plus or minus'' property (``plus'', in this
case), then it cannot have a definite value of the complementary
property ``dead or alive''.

\section{Complementarity and the Bell theorem}
One may be tempted to dismiss quantum complementarity by stating that
a cat in the state $(|dead\>+|alive\>)/\sqrt{2}$ has also a definite
value of the ``dead or alive'' property but, somehow, we are ignorant
of it. Namely, we can argue that complementarity is not a
limitation on the values of the properties that an object can possess,
but rather on the fact that a quantum measurement can extract only one
of them at a time (and, possibly, perturbs the values of the
complementary ones).

In order to maintain such position, however, one needs to give up
Einstein locality (technically the ``no signaling
condition''). Indeed, Bell's theorem \cite{bell} tells us that quantum
mechanics is incompatible with local hidden (i.e.~unknown) values. A
simple exposition of Bell's theorem and its implications is in
\cite{belllor}. This means that, if one wants to retain Einstein
locality, values of complementary properties are not even defined in a
quantum system. One cannot say that they are defined, but unknown. Of
course, one could retain unknown values at the cost of locality, as is
done in Bohmian mechanics, where the hidden variables are notoriously
nonlocal (signaling)\footnote{At the cost of stating the obvious,
  quantum mechanics {\em is} nonlocal in the {\em quantum sense}
  (i.e.~it contains nonlocal correlations), but it is an
  interpretation-dependent statement whether it is nonlocal in the
  {\em Einstein sense} \cite{belllor}: Bohmian hidden variables are
  nonlocal in the Einstein sense (their knowledge would allow
  signaling), but they are absent in the Copenhagen interpretation
  which can then be considered local in the Einstein sense
  \cite{copenhagenislocal1,copenhagenislocal2,copenhagenislocal3,copenhagenislocal4},
  thanks to complementarity.}. In this paper we take the point of view
of Copenhagen quantum mechanics.

In fact, by measuring complementary observables on entangled states,
one can observe \cite{exp1,exp2,exp3} correlations among the
measurement outcomes which are incompatible (Bell's theorem) with any
possible prescription that assigns pre-determined values to these
properties, unless one postulates that the measurement of one property
somehow is able to nonlocally change the value of a property of the
correlated distant system. Most physicists (not all!) are unwilling to
abandon Einstein locality, because of the causality problems that
would emerge if one could access these hypothetical
``instantaneously''\footnote{This word is meaningless unless one
  specifies which reference frame it refers to.} \cite{gisin}  propagating hidden
variables and because one would have to introduce a preferred
reference frame which would violate the postulate of relativity that all
inertial frames are equivalent. Then Bell's theorem forces us to admit
that if one observable is well defined, then the values of
complementary ones are not even defined. I emphasize that
``undefined'' is very different from ``unknown''. While an undefined
quantity is clearly also unknown, the opposite is not true. For
Copenhagen, the state \eqref{cats} represents a state where the
``being dead or alive'' property is undefined, whereas the state
\eqref{rhocat} represents a state where it is
unknown\footnote{Historically, at the time the cat paper was
  published, Einstein \cite{epr} and Schr\"odinger \cite{cat} were
  convinced that quantum mechanics was incomplete. Einstein's stance
  referred to the fact that the quantum state should only refer to
  information about an ensemble of systems, and not about a single
  system \cite{ballentine,ballentine2}. He apparently never proposed
  that deterministic hidden variables should be used to describe
  single systems, and criticized Bohm for doing so
  \cite{ballentine}. Schr\"odinger's stance (\cite{cat}, Sec.13) seems
  to veer in the direction that hidden variables (not present in
  Copenhagen's interpretation) might keep track of the values of
  complementary observables. Bohr, of course, dissented
  \cite{bohrcom}. Bell's theorem (formulated after Einstein's and
  Schr\"odinger's deaths) showed that any hidden variables that lead
  to the statistical predictions of quantum mechanics would violate
  Einstein's locality, arguably a vindication of Bohr's point of
  view.}.

Then the Schr\"odinger cat, thanks to complementarity, has an
undefined value of whether it is alive or dead. A quite weird
situation for the poor cat!
\section{Public outreach}
In this section we detail how one can create a simple Arduino-based
simulation of the cat experiment using a cardboard box.

\begin{figure}[hbt]
\begin{center}
  a)
\epsfxsize=.9\hsize\leavevmode\epsffile{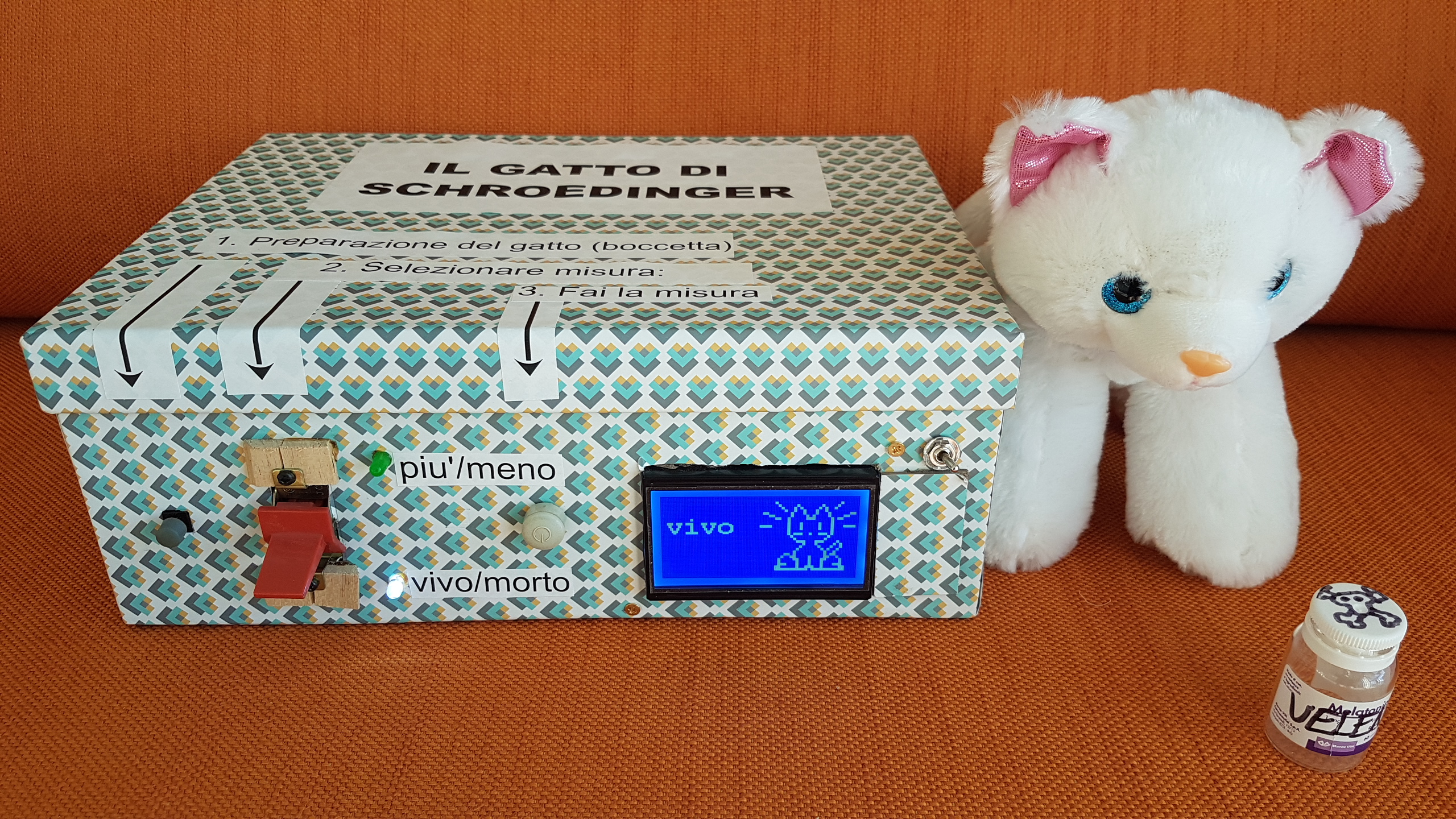}
\end{center}
\begin{center}
  b)
\epsfxsize=.9\hsize\leavevmode\epsffile{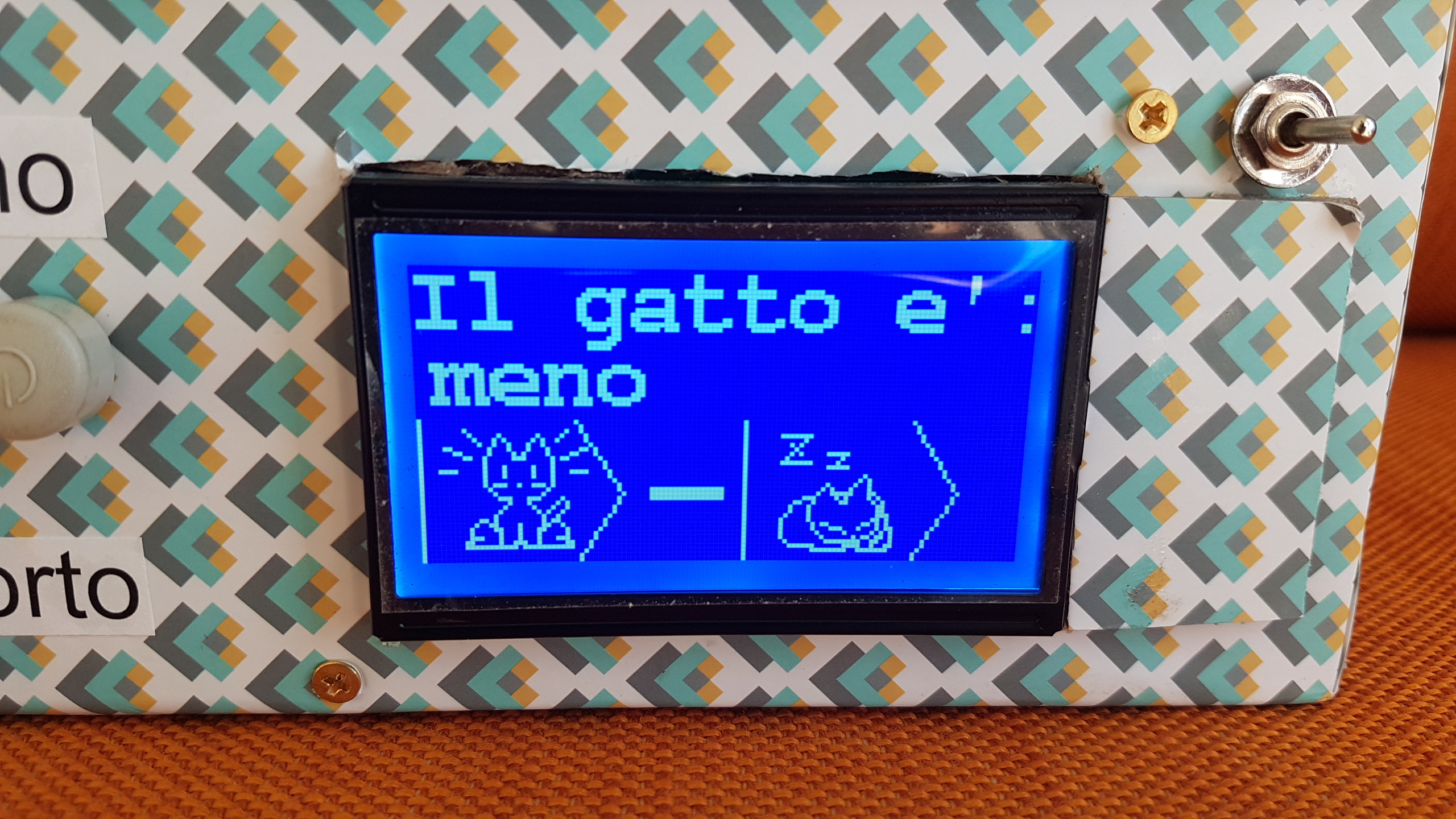}
\end{center}
\vspace{-.5cm}
\caption{a). Photo of the Schr\"odinger cat simulated experiment for
  science outreach. A stuffed cat and a small (simulated) poison
  vial are good props to keep people's attention. b) Close-up of the
  display. Here it is showing the post-measurement state after the
  measurement of $\hat S$ has given outcome $-1$. See text for the
  description.} \labell{f:scatola}\end{figure}

On the box front (see Fig.~\ref{f:scatola}) there are a series of
input switches. On the left a press-button switch simulates the
preparation of the cat state, namely the activation of Schr\"odinger's
infernal machine. After pressing it, the display announces that the
cat is in the ``plus'' state, the one described by the state
\eqref{cats} relative to the eigenvalue $+1$ of the $\hat S$
operator.

Then, using a selector switch to the right of the press-button, the
``experimenter'' can choose which of the two complementary properties
to measure: either the ``dead/alive'' or the ``plus/minus''
property. The first refers to the measurement of the cat in the
$|dead\>$ and $|alive\>$ basis, the second is a measurement of
$\hat S$, namely the $(|dead\>+|alive\>)/\sqrt{2}$ and
$(|dead\>-|alive\>)/\sqrt{2}$ basis. A led lights up to confirm the
choice of measurement: a green led for the $\hat S$ measurement and a
white led for the dead/alive measurement.

Finally, a last press-button to the right simulates the activation of
the measurement of the previously chosen property. The display returns
the measurement outcome. If the cat state is in one of the eigenstates
of the chosen measurement, then that is the outcome. E.g.~if the cat
is in an $|alive\>$ state and the ``dead/alive'' measurement is
selected, the outcome will be ``alive''. Otherwise, the outcome is
chosen at random with uniform distribution and is the cat's state is
updated to the eigenstate relative to the obtained outcome (collapse
of the state).

After the outcome is presented, the display shows the current state of
the cat, and one can perform a new measurement or one can reprepare
the cat in the state \eqref{cats} by pressing the first button.

A switch-sensor determines whether the box is opened. In this case,
the measured property (indicated by the led) is automatically switched
to the ``dead/alive'' property, and the cat is measured in the
$|dead\>$ or $|alive\>$ basis.

Clearly, this is not a historically accurate reproduction of
Schr\"odinger's proposal, since he never advocated for complementary
properties. Nonetheless, the goal is science outreach: explaining in a
clear fashion quantum complementarity and superposition, rather than
historical accuracy.

The box is powered by an Arduino nano microcontroller (but any Arduino
variant will work). The instructions to build it, the software and an
illustrative movie of its operation can be found here
\cite{nextcloud}. The cost is about 25-30 dollars (or euros),
excluding the power source. It can be powered through any powerbank or
usb charger. An approximate bill of materials of the main components:
Arduino nano: $4\$$; LCD display ST7920: $8\$$; cardboard box: $5\$$;
stuffed cat: $8\$$.  Since the box is intended for an Italian public,
all labels on the box and messages on the display are in Italian, but
they can be readily translated to any local language.

\section{Conclusions}
In conclusion, we have presented a simple interpretation of
Schr\"odinger's cat, based on quantum complementarity. The superposed
cat has a definite value of a property $\hat S$ with eigenstates
$(|dead\>\pm|alive\>)/\sqrt{2}$ which is, hence, {\em complementary}
to the property ``being dead or alive''. Such interpretation uses only
standard textbook concepts. A simulated Schr\"odinger cat experiment
is presented, together with the indications of how to cheaply
replicate it. It has been tested in several science outreach
occasions.

\section{Acknowledgments}
I acknowledge useful feedback from the organizers of the Italian
Quantum Weeks (IQW), especially Maria Bondani, and from Claudio
Sutrini. I also acknowledge an interesting correspondence with Arkady
Plotnisky. This work received support from EU H2020 QuantERA ERA-NET
Cofund in Quantum Technologies, Quantum Information and Communication
with High-dimensional Encoding (QuICHE) under Grant Agreement 731473
and 101017733, from the U.S. Department of Energy, Office of Science,
National Quantum Information Science Research Centers, Superconducting
Quantum Materials and Systems Center (SQMS) under Contract
No. DE-AC02-07CH11359, from the PNRR MUR Project PE0000023-NQSTI, from
the National Research Centre for HPC, Big Data and Quantum Computing,
PNRR MUR Project CN0000013-ICSC.

\end{document}